\documentclass[aps,prd,onecolumn,superscriptaddress,showpacs,floatfix,nofootinbib]{revtex4}
\usepackage{graphicx}
\begin{document}

%
%

\title{Particle identification for Higgs physics at 
future electron-positron collider}

\author{Wei-Ming Yao\footnote{Presented at HKUST IAS Program High Energy Physics, Hong Kong, January 7-25, 2019}}

\address{Lawrance Berkeley National Lab, MS-50B-5239 \\
One Cyclotron Rd, Berkeley, California 94720, US \\
wmyao@lbl.gov} 



\begin{abstract}
Particle Identification (PID) plays a key role in heavy flavor physics
in high-energy physics experiments. However, its impact on Higgs physics
is still not clear. In this note, we will explore some
of the potential of PID to improve the identification of heavy-flavour jets 
by using identified charged Kaons in addition to the traditional vertexing
information. This could result in a better measurment of the Higgs-charm Yukawa coupling
at the future e+e- colliders.
\keywords{Particle-identification; Higgs; Heavy-flavour tagging; Silicon-tracker; Future-collider}
\end{abstract}

\maketitle  



\section{Introduction}	

The observation of the Higgs boson (H) by ATLAS and CMS in July 
2012~\cite{hatlas,hcms} marked the completion of the standard model (SM) 
and opened the way to explore the Higgs sector that is responsible for 
electroweak symmetry breaking (EWSB). The HL-LHC will probe the Higgs
physics at an unprecedented level. The planned future lepton collider 
(ILC, CEPC or FCC-ee) seems to be the next logical machine~\cite{ilc,cepc,fccee}. By running at 
multiple center of mass energies at Z-pole,
$W^+W-$, $ZH$, and possible $t\bar t$ thresholds the new machine will allow the Higgs coupling to 
be measured precisely at or below a percent level,  as well as to constrain 
the BSM physics via direct search and precision electro-weak data.

In order to achieve these physics goals, the proposed detector concept   
must meet the stringent performance requirements in a large solid-angle 
coverage for excellent particle identification, 
precise particle energy/momentum measurement, efficient vertex 
reconstruction, and superb jet reconstruction and measurement, as well as 
the flavor tagging. One of the tracking concepts is to use a 
full silicon tracker (FST), which consists of a pixel vertexing detector and 
a double-sided silicon strip detector. It has excellent spatial resolution and 
granularity to separate tracks in the dense jets and high 
occupancy from beam-related backgrounds at high luminosities, but on the other hand, 
it has limited dE/dX resolution for particle identification (PID) compared to 
other proposed tracking options such as TPC or draft chamber~\cite{cepc,fccee}. 

In this short note, we explore some of the potential of PID
to improve identification for charge and flavor of the heavy-flavour jets. 
The identified charged Kaon particles inside the jets are used to 
improve the discrimination between b- and c- jets, which could result in a better measured value for 
the Higgs-charm Yukawa coupling at the future e+e- collider. 
 
\section{Case study for PID} 
\begin{figure}[htpb]
\centerline{
\includegraphics[width=3.8cm]{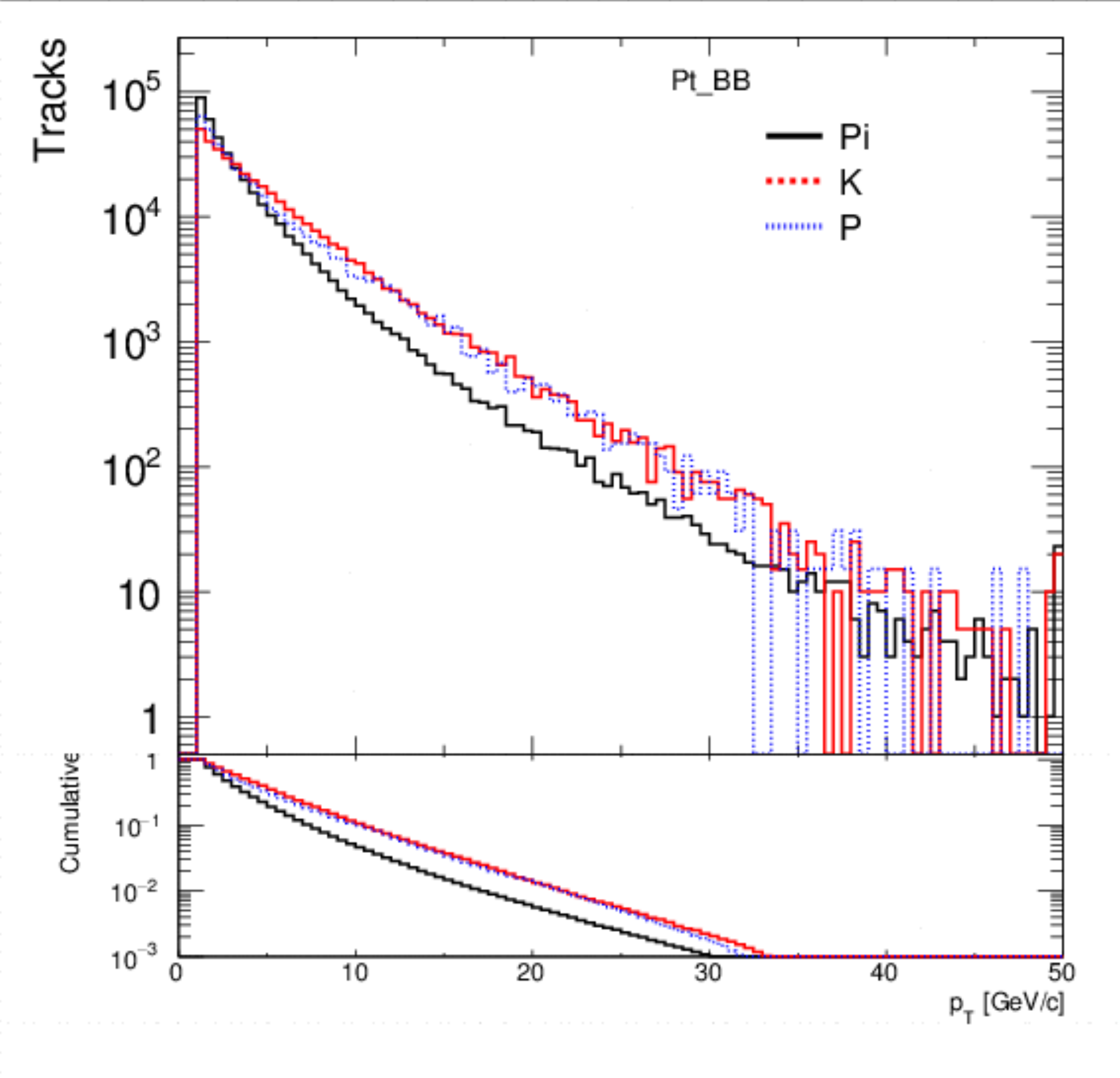}
\includegraphics[width=3.8cm]{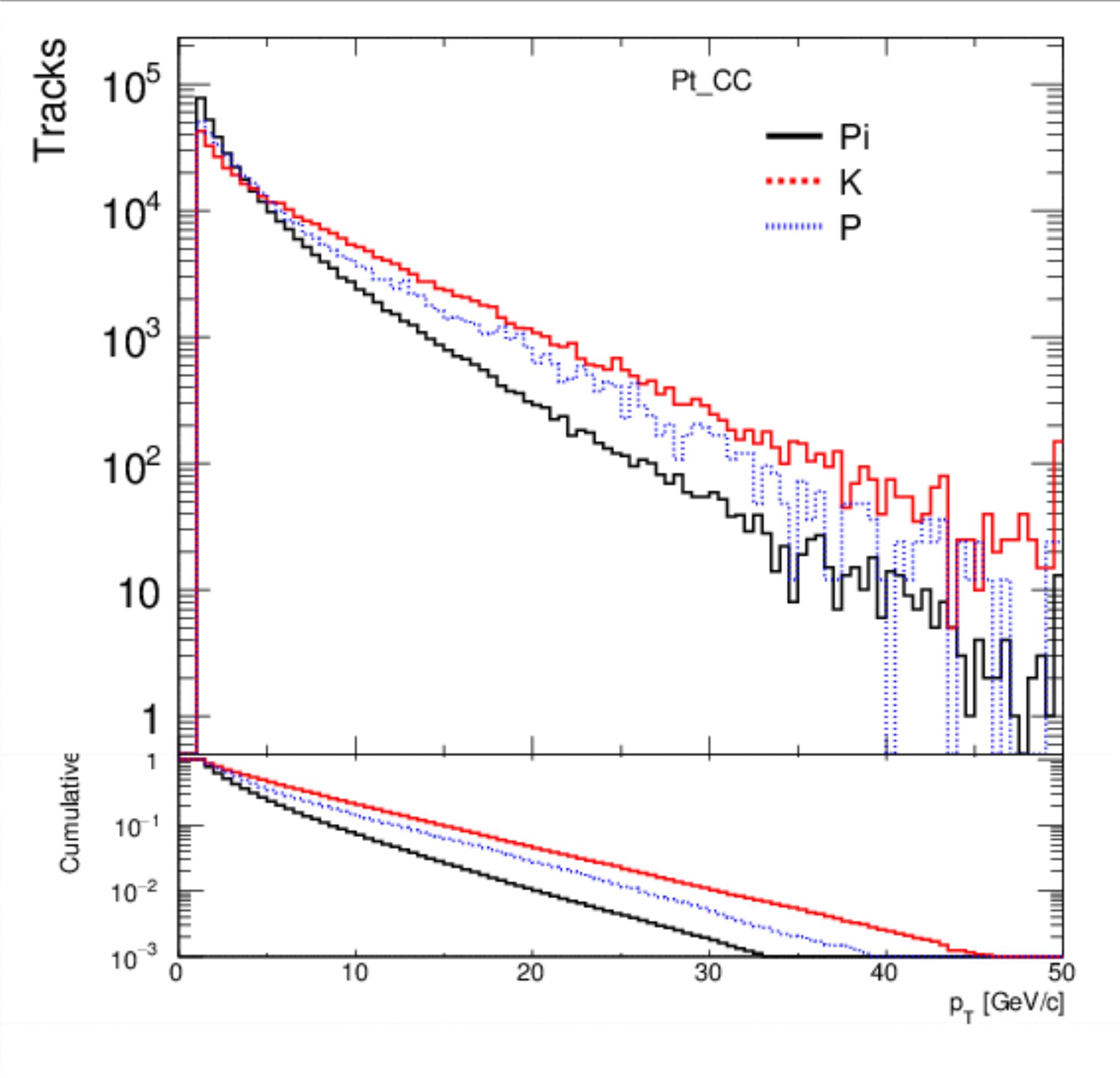}
\includegraphics[width=3.8cm]{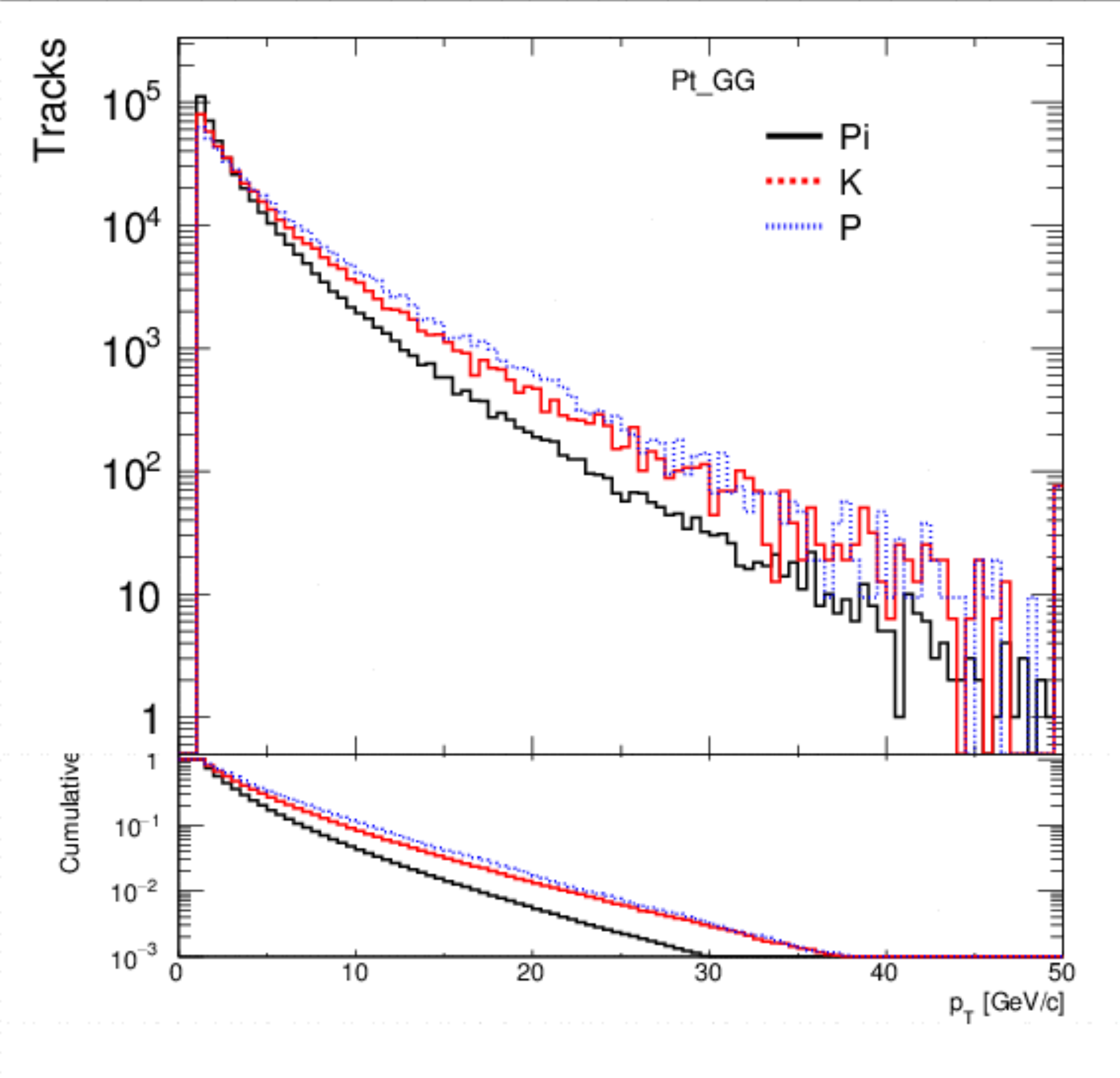}
\includegraphics[width=3.8cm]{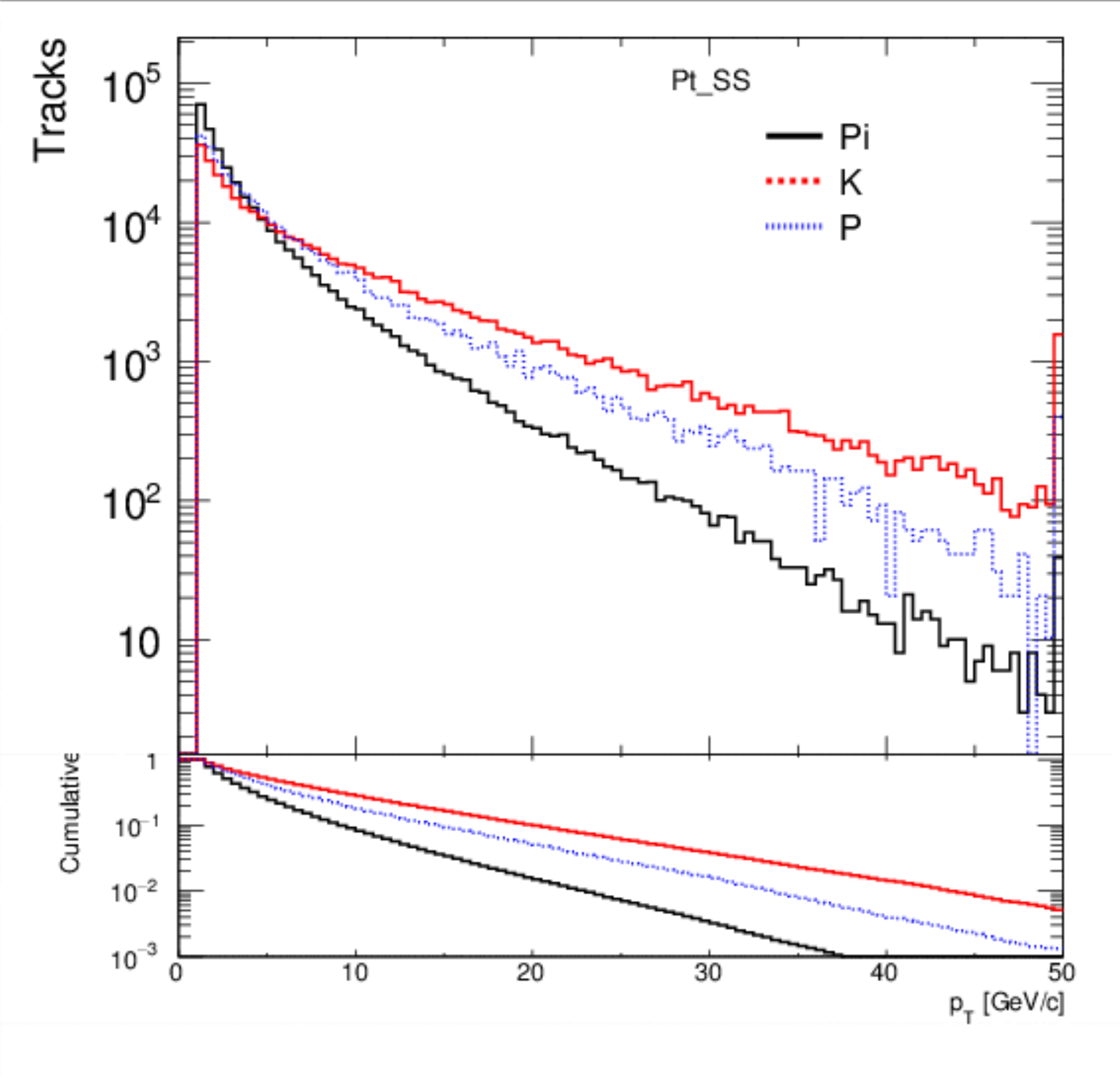}}
\caption{The track $P_t$ distribution for pion, kaon, and proton from
$H\rightarrow B\bar B$, $C\bar C$, $GG$, and $S\bar S$ and
their accumulative fractions.\label{fig:trackpt}}
\end{figure}

\begin{figure}[htpb]
\centerline{
\includegraphics[width=3.8cm]{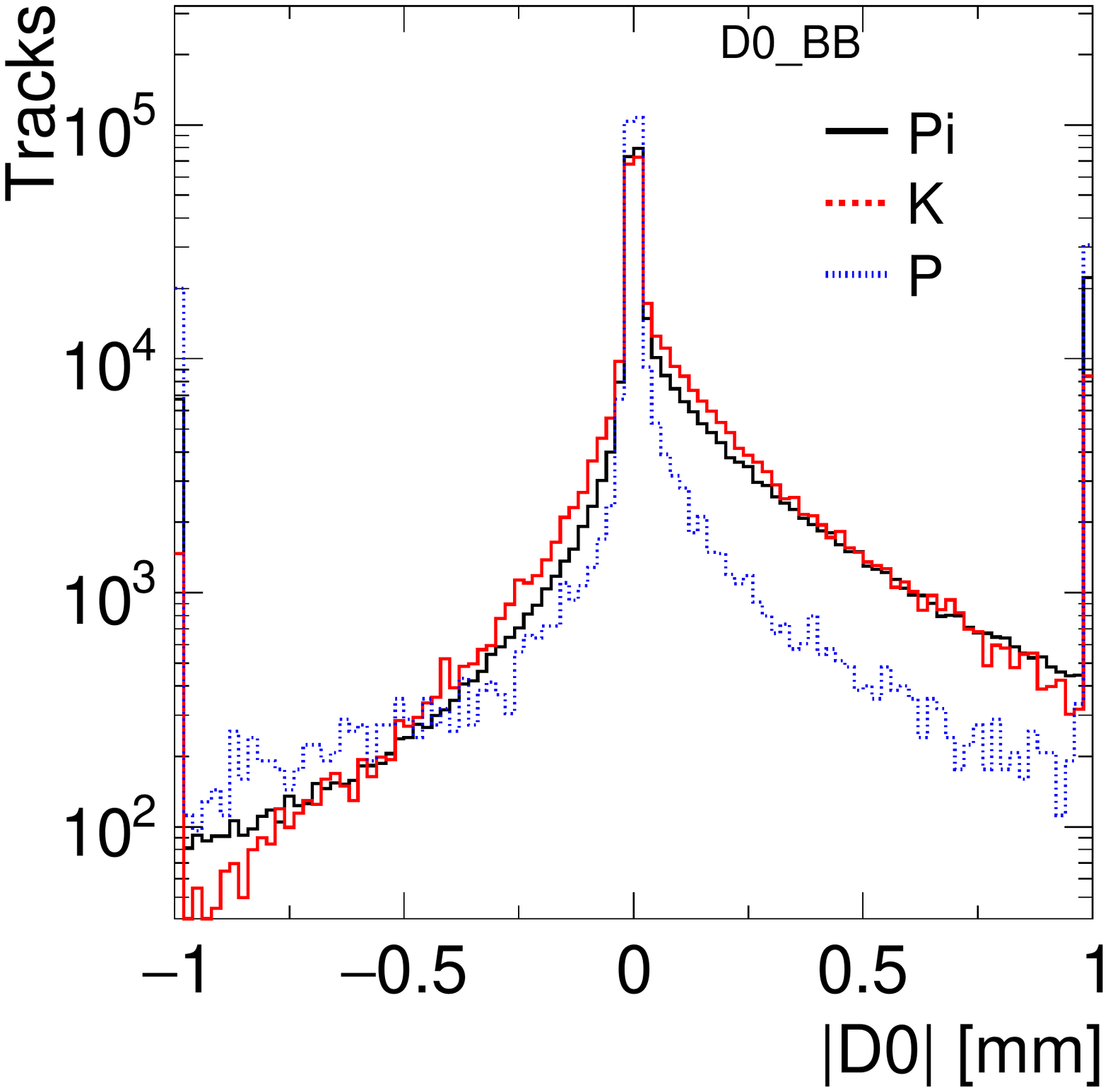}
\includegraphics[width=3.8cm]{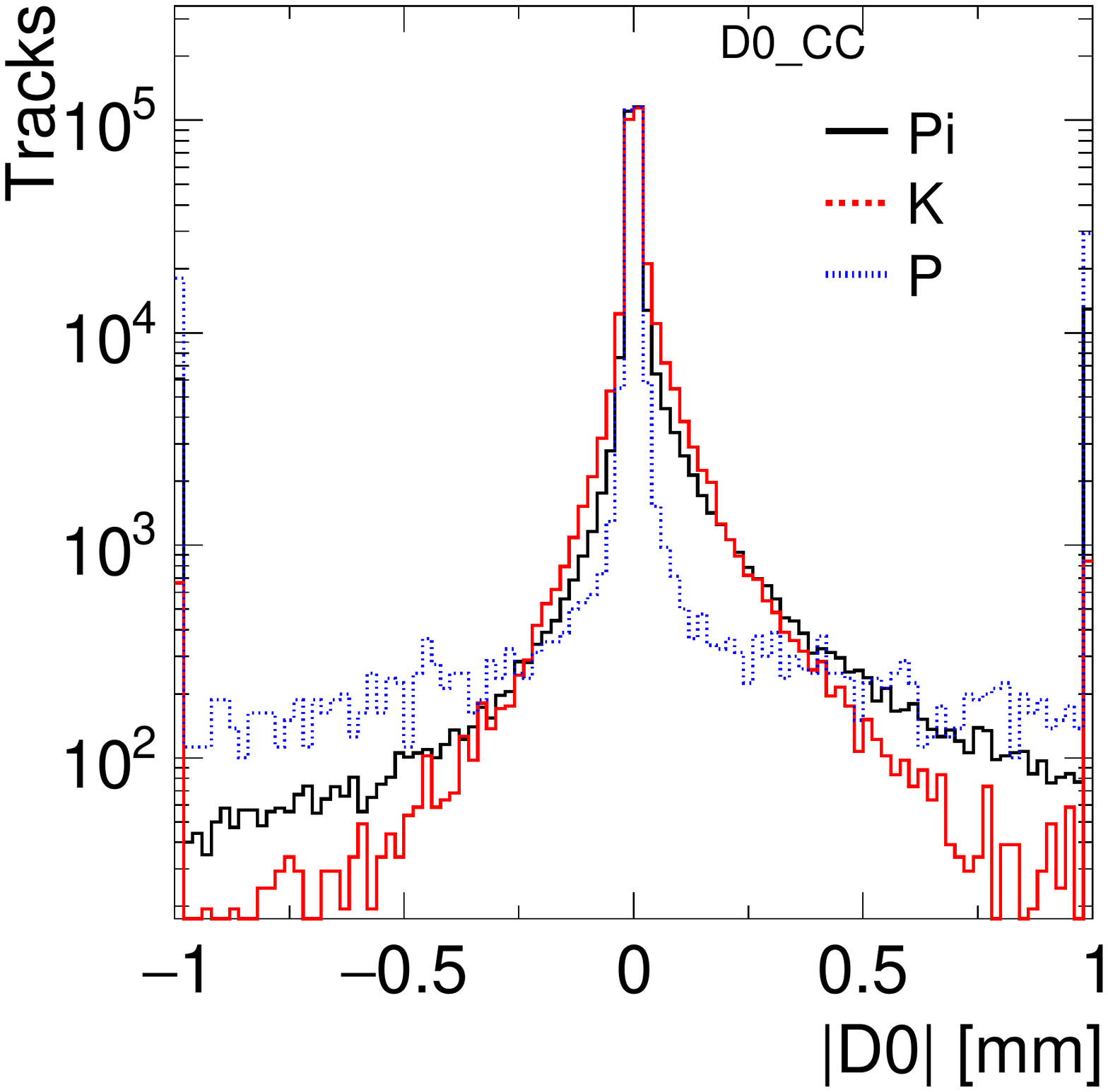}
\includegraphics[width=3.8cm]{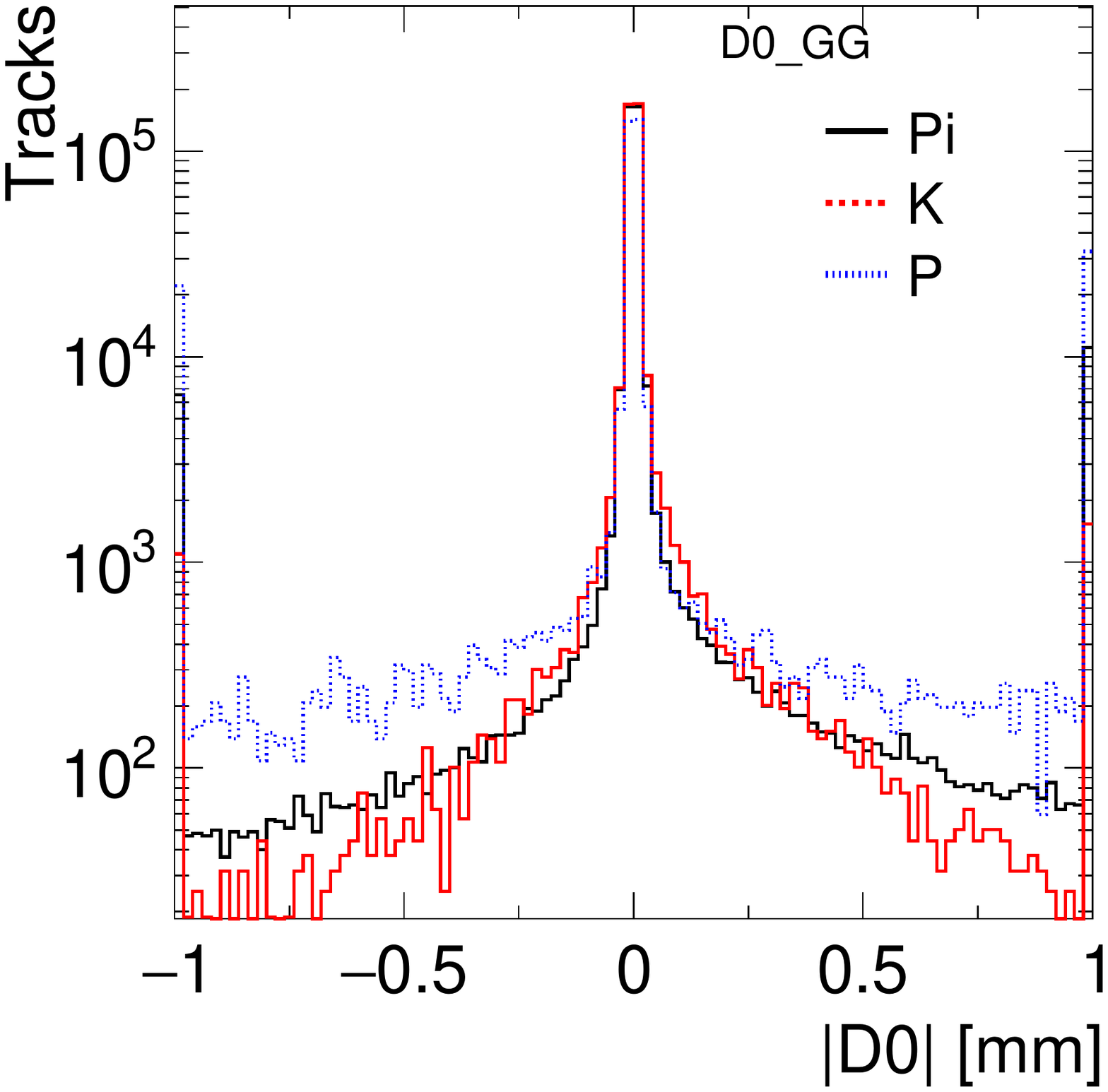}
\includegraphics[width=3.8cm]{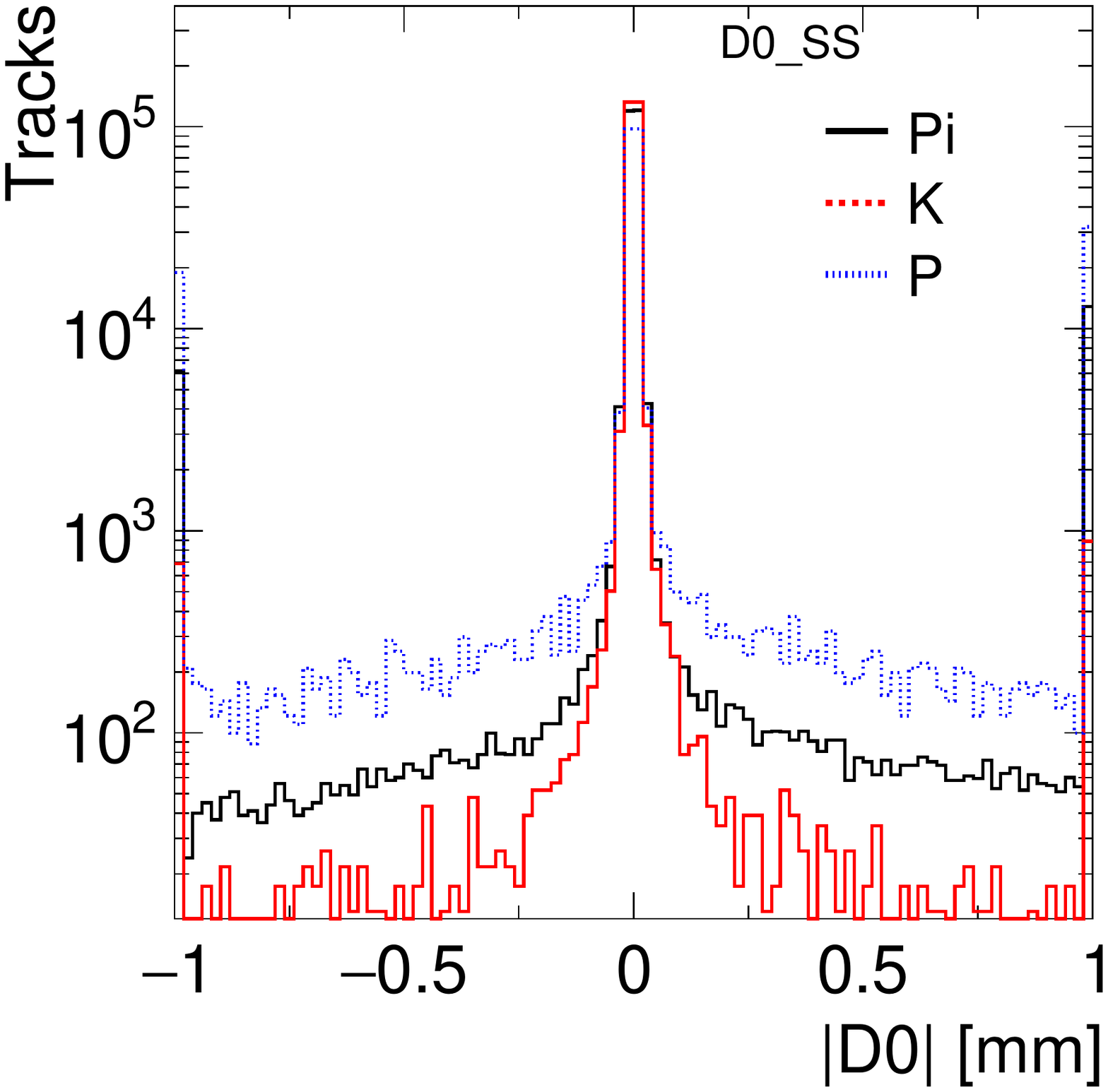}}
\caption{The track signed $D_0$ distributions for pion, kaon, and proton from 
$H\rightarrow B\bar B$, $C\bar C$, $GG$, and $S\bar S$.\label{fig:sd0}}
\end{figure}

We used several ZH samples generated using CEPC v4 detector configuration where
$Z\rightarrow \mu^+\mu^-$ and the Higgs boson decays into $b\bar b$, $c\bar c$,
$gg$, and $s\bar s$~\cite{cepcsf}. The jets are reclustered using anti-kt algorithm with 
the particle flow objects after removing the muon tracks from Z decay. Each event is 
forced to have a di-jet topology. The tracks inside the 
jets are selected after matching to the Monte Carlo truth. The charged particle Pt are shown in Fig.~\ref{fig:trackpt}
for selected Pion, Kaon and Proton tracks from  $H\rightarrow 
b\bar b$, $c\bar c$, $gg$, and $s\bar s$, respectively.  The integrated 
cumulative distribution is also shown at the bottom of plot. Most tracks are $P_T<30$ GeV/c. 
The signed impact parameter of these tracks (sd0) with respect to the jet direction
is also shown in Fig.~\ref{fig:sd0}. The sign is defined as $cos(\phi_{trk} + \pi/2 * sign(d0)-\phi_{jet})$ where 
$\phi_{trk}$, $\phi_{jet}$, and d0 are the azimuth angles of track and jet, and the impact parameter, respectively. 

\begin{figure}[htpb]
\centerline{
\includegraphics[width=3.8cm]{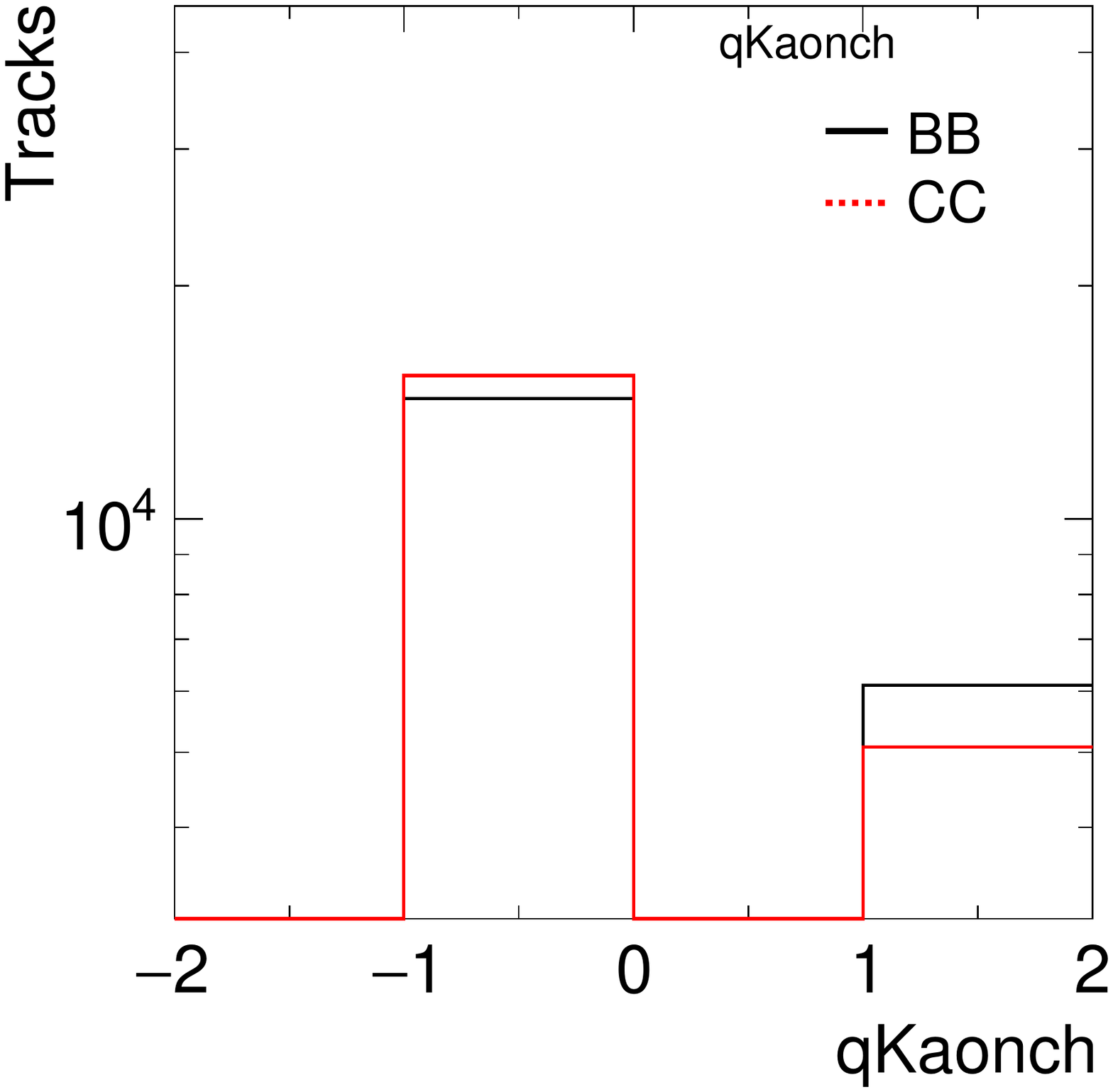}
\includegraphics[width=3.8cm]{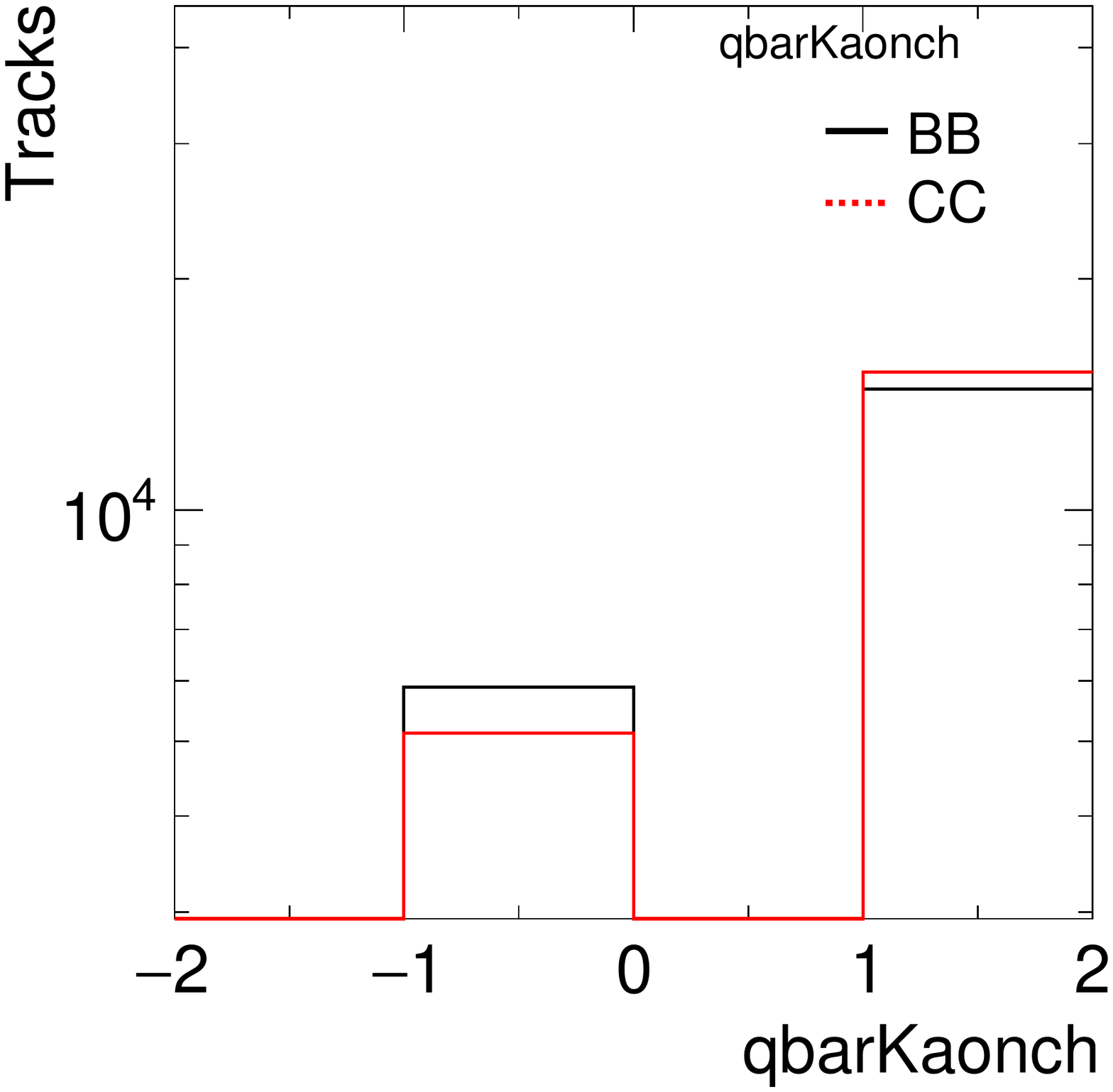}}
\caption{The leading kaon charge for $b/c$-jet and for $\bar b/\bar c$-jet.\label{fig:charge}}
\end{figure}

To illustrate the importance of PID, we select charged Kaon particles inside the jets based on their MC truth information. 
The charged Kaon is strongly correlated with the charge of its parent quark. 
The b and c quarks decay into $K^-$. On the other hand, the $\bar b$ and $\bar c$ quarks decay into $K^+$. 
The correlation between the charge of the leading 
Kaon and the charge of the orignal quark is shown in Fig.~\ref{fig:charge}, which gives a correlation of 70\%.
The jet charge identification can be further improved by including vertexing and additional lepton tracks, which can be used 
to improve jet-pairing in the construction of $H\rightarrow VV\rightarrow 4j$ in the all-haronic final states,
where V stands for W or Z boson decaying into di-jets. 

\begin{figure}[htpb]
\centerline{
\includegraphics[width=3.8cm]{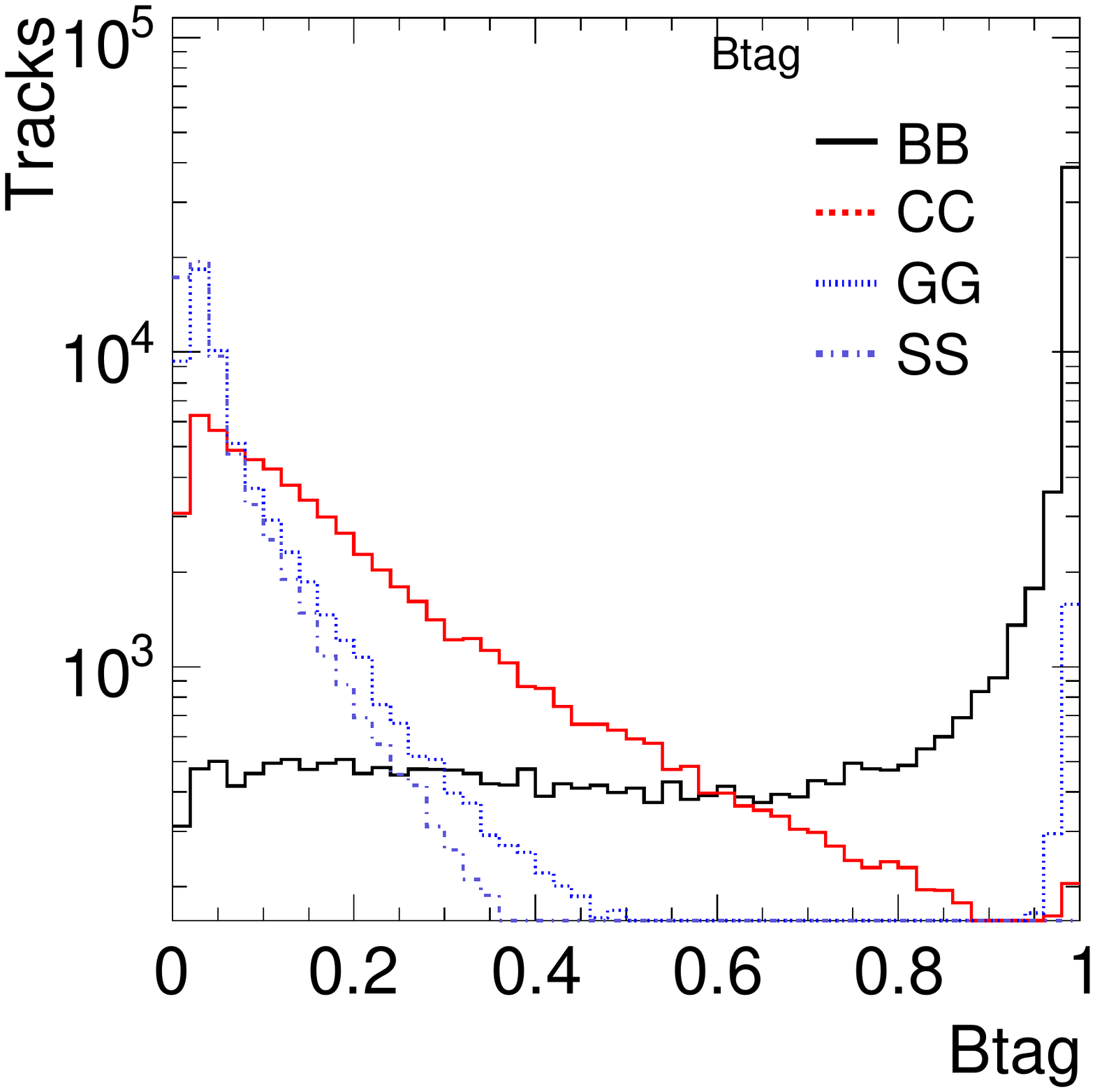}
\includegraphics[width=3.8cm]{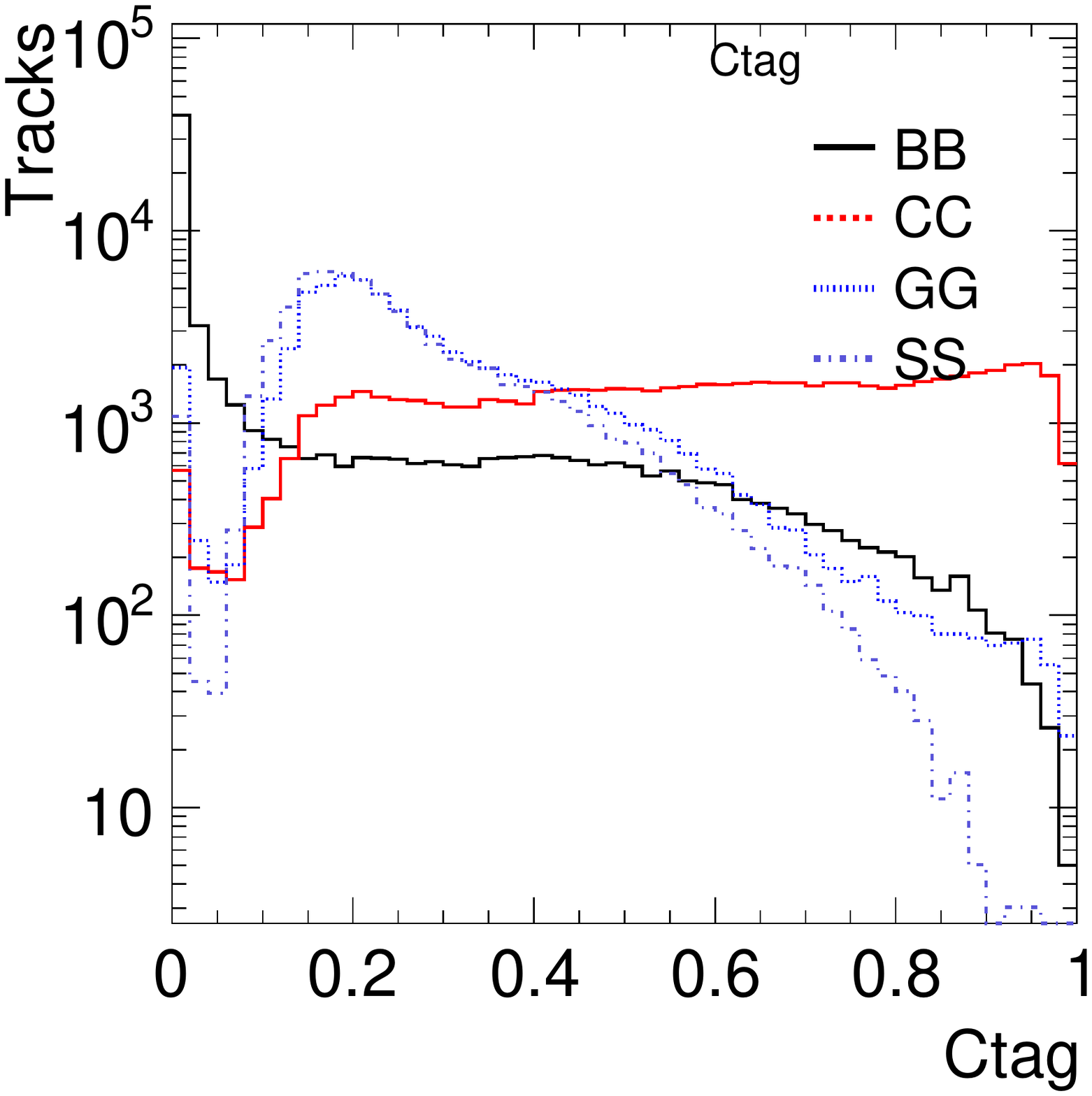}}
\caption{The standard btag output for 
track $P_t$ distribution for pion, kaon, and proton from
$H\rightarrow B\bar B$, $C\bar C$, $GG$, and $S\bar S$ and
their accumulative fractions.\label{fig:tag}}
\end{figure}

The PID can also be used to improve the charm jets tagging by identifying the 
charged Kaon tracks inside the jet. The dominated background for charm tagging is due to  
the contamination of b-jet. The $b$- and $c$-jets are currently 
identified by ILCbtag using a BDT trained on the displaced tracks~\cite{cepcsf}. 
The output of BDT for the $b$- and $c$-jets are shown in Fig.~\ref{fig:tag} as $B_{tag}$ and 
$C_{tag}$. We select a charm enriched sample with $C_{tag}>0.2$ to test the idea whether the PID is useful 
for improving the existing charm tagging. 
A new BDT is trained using the identified charged Kaon tracks and
the displaced tracks with a significance of the signed impact parameter ($sd0>2.0$) inside the jets, as well as 
other variables listed below:
\begin{itemize} 
\item LeadkPt: the $P_t$ for identified the leading charged Kaon in GeV. 
\item Summed charges: the summed charge of the leading Kaon and displaced tracks.
\item Mass: the invariant mass of the leading Kaon and displaced tracks.
\item SumPt: the summed $P_T$ of the leading Kaon and displaced tracks.
\end{itemize}    

\begin{figure}[htpb]
\centerline{
\includegraphics[width=3.8cm]{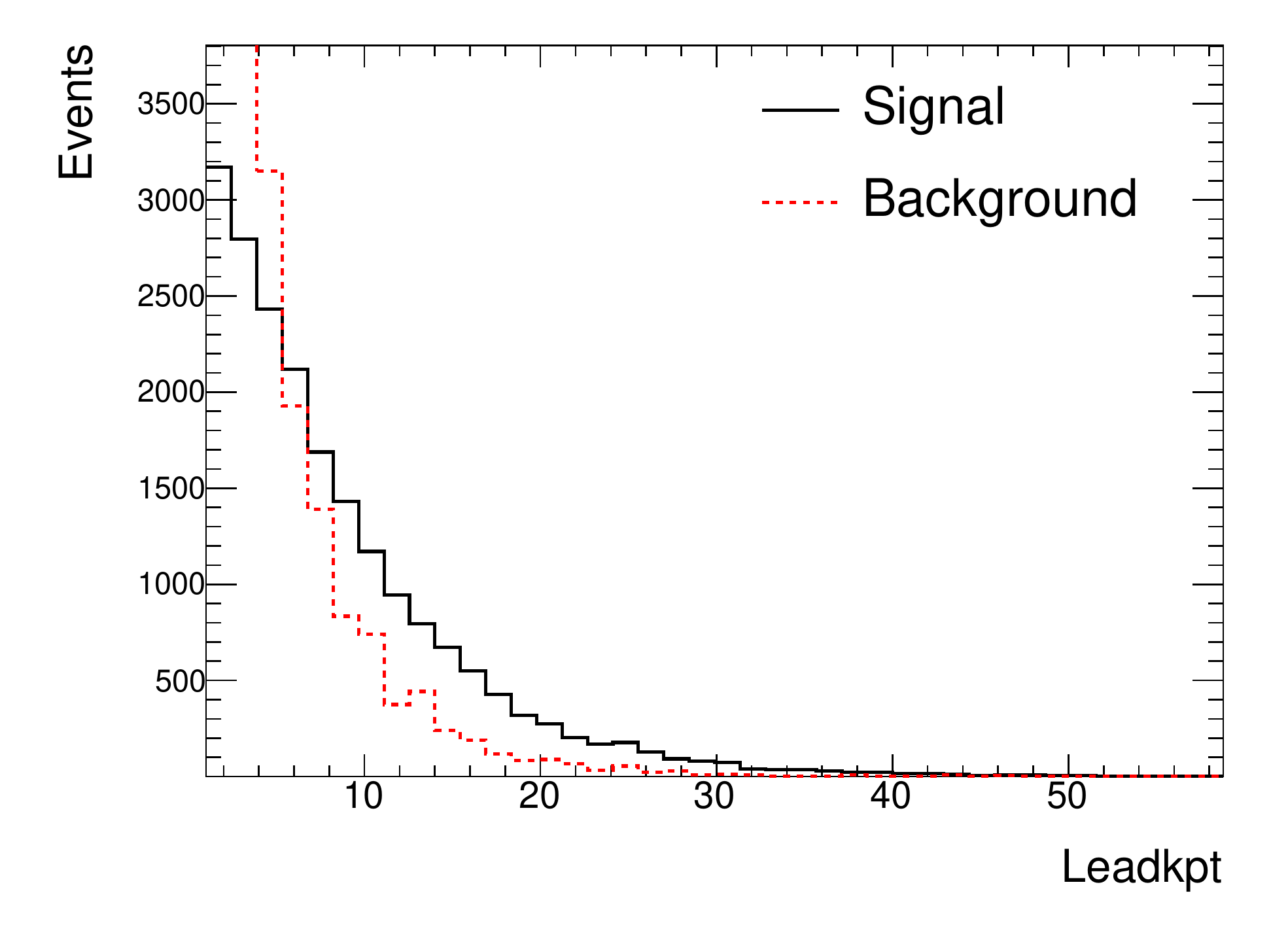}
\includegraphics[width=3.8cm]{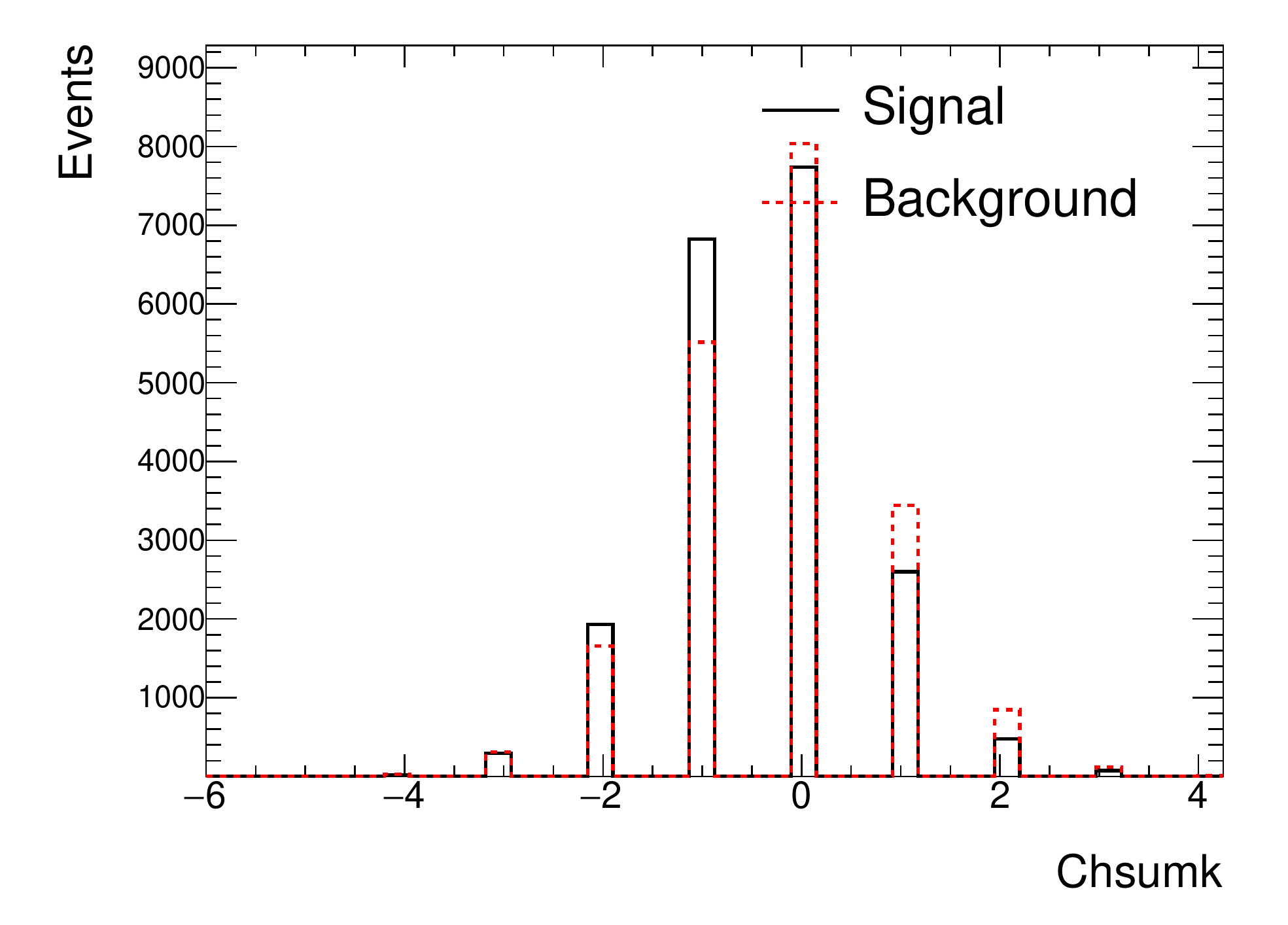}
\includegraphics[width=3.8cm]{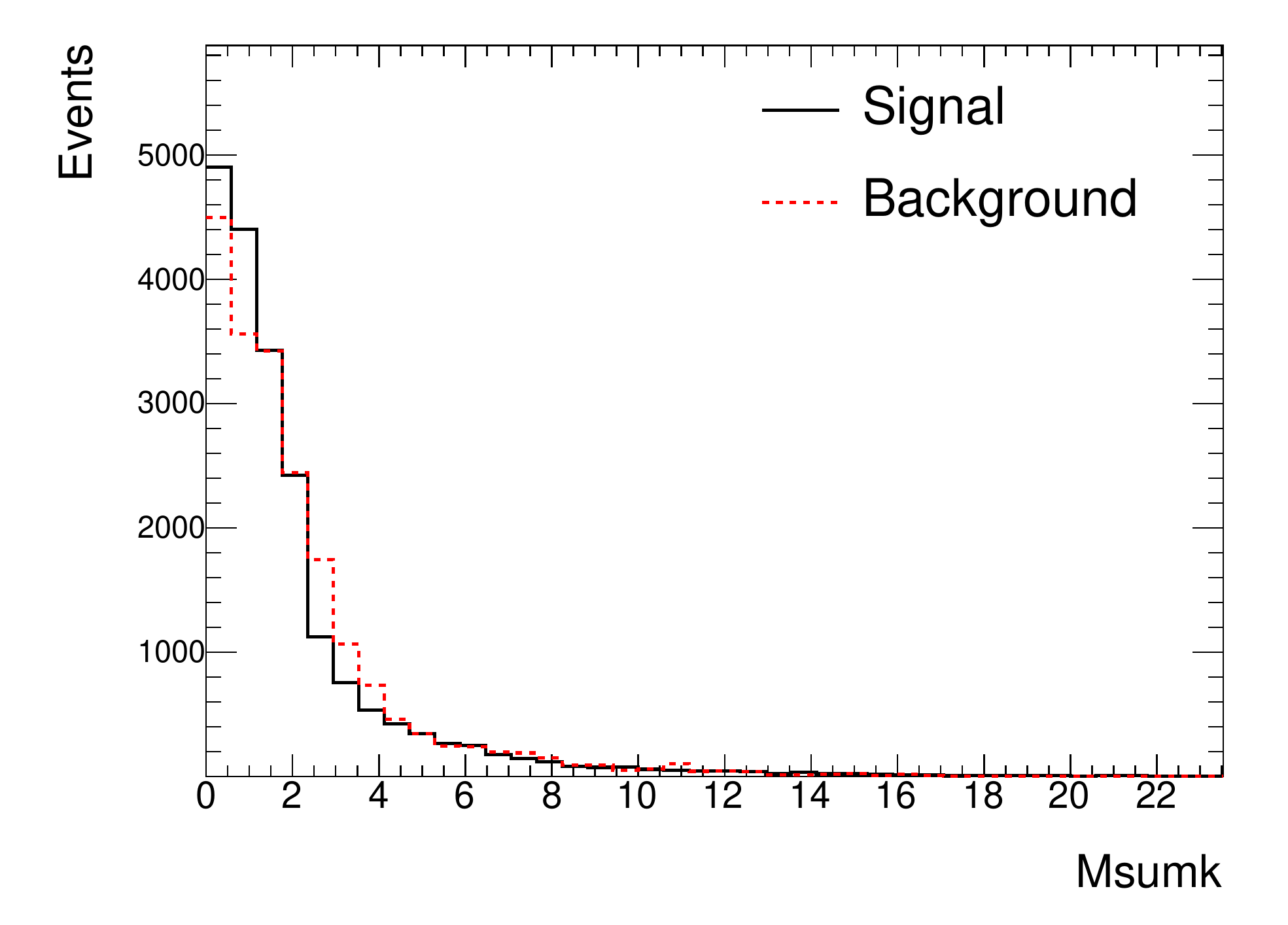}
\includegraphics[width=3.8cm]{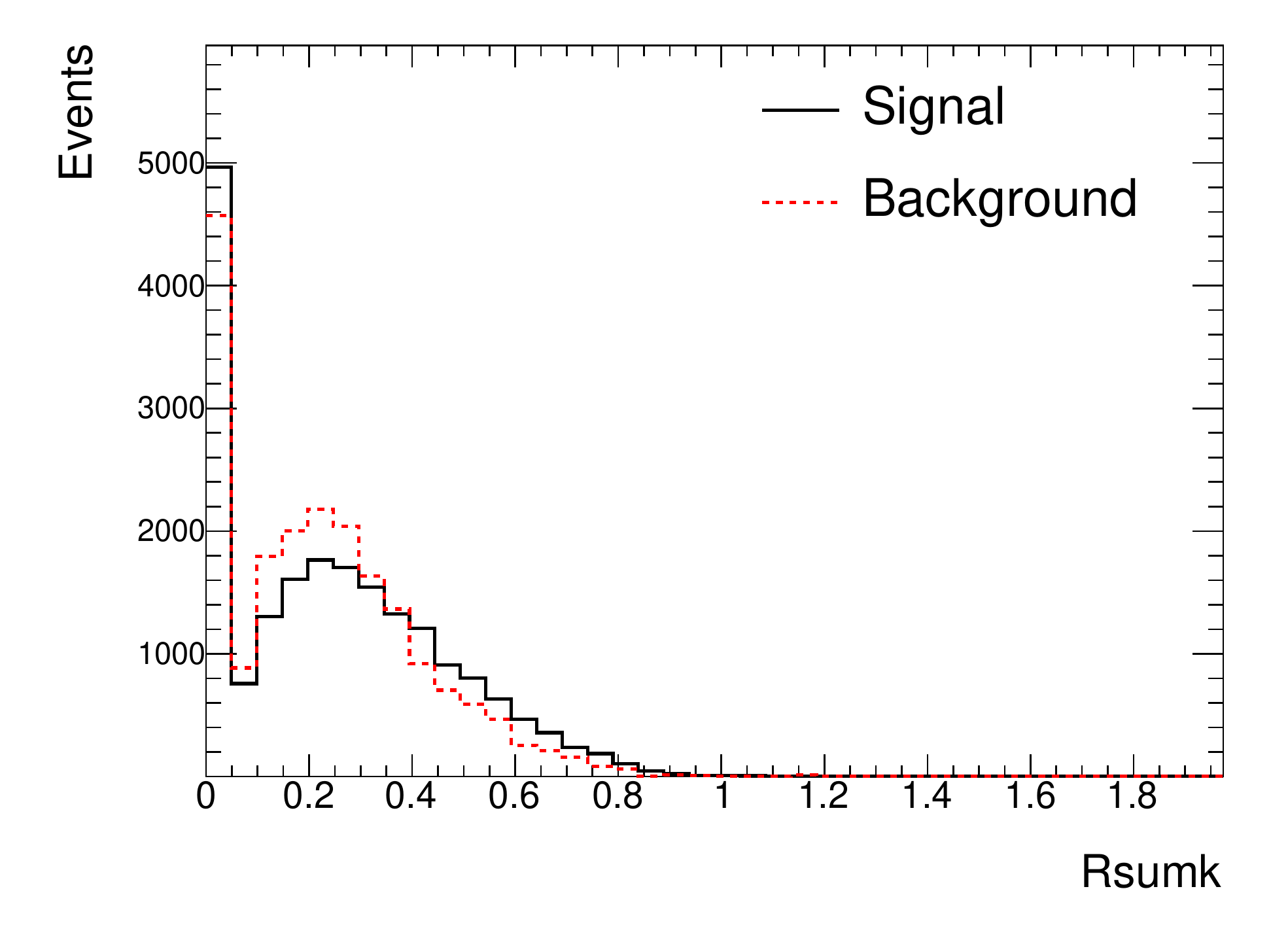}}
\caption{The input variables are compared between $H\rightarrow c \bar c$ and $H\rightarrow b\bar b$ 
for leading kaon $P-T$, the summed charge, the invariant mass, and the summed $P-T$ of the leading kaon and displaced tracks,
respectively.\label{fig:newinputs}}
\end{figure}

\begin{figure}[htpb]
\centerline{
\includegraphics[width=3.8cm]{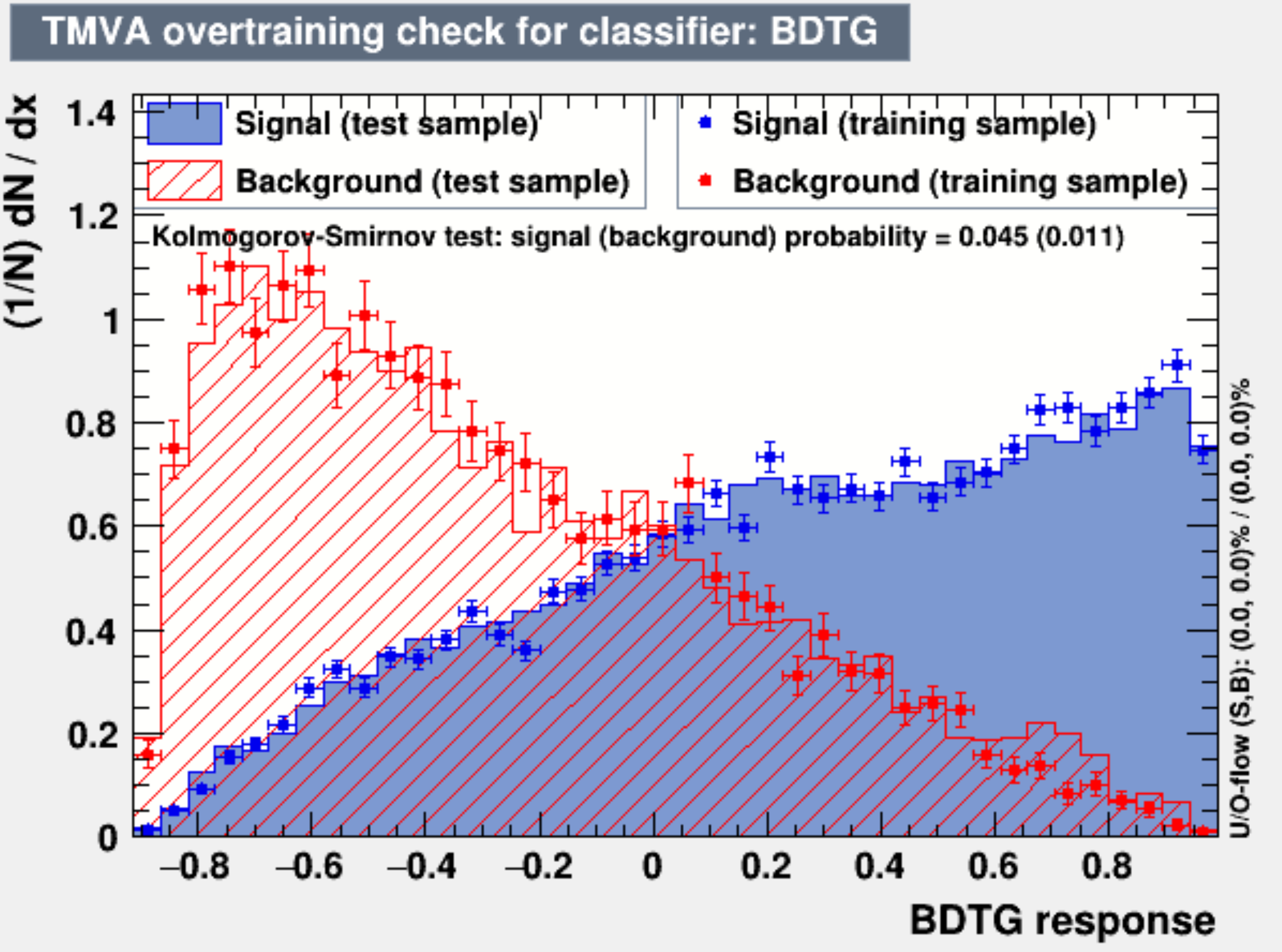}
\includegraphics[width=3.8cm]{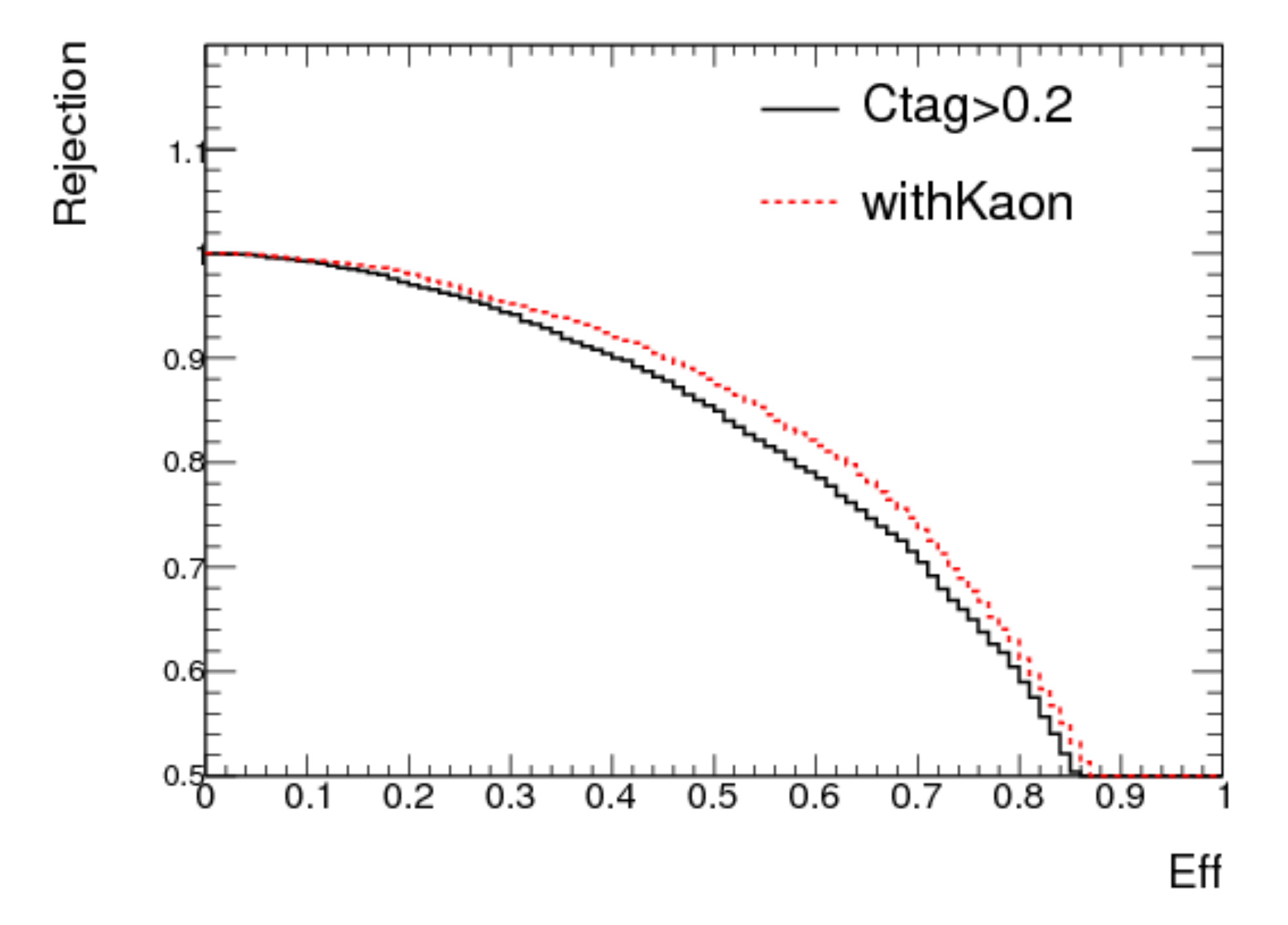}}
\caption{The over-trained BDT and the ROC curve are shown with and without additional kaon variables.\label{fig:newbdt}}
\end{figure}

The comparison of distributions from the $c$-jet(signal) and the $b$-jet
(background) are shown in Fig.~\ref{fig:newinputs}. The new trained BDT and the ROC curve are 
shown in Fig.~\ref{fig:newbdt}. The improvement is about 10\% in terms of rejection 
while keeping the same efficiency. Further improvements are possible by 
seeding the vertex reconstruction using the identified charged Kaon particles.
The $c$-jet usually contains one secondary vertex from 
$c\rightarrow s$ decay while the $b$-jet contains two secondary vertices from $b\rightarrow c$ and 
$c\rightarrow s$ decays. The identified leptons from $b$- or $c$ decay are also useful to separate the $b$- 
and $c$-jet as well. 
 
\section{PID options for full silicon tracker}
We present a proposal for using fast timing and the Ring Image of Cherenkov (RICH) detectors to improve the particle identification capabilities 
for the full silicon tracker detector option for future $e^+e^-$ 
colliders. 
The last two outer-strip layers can be replaced with the high
granularity fast silicon photomultipliers detectors (SiPM)~\cite{lgad, ufsd}. 
which can be used to detect both charged tracks and a single photon 
with a timing resolution of $\approx$ 30~ps and a spatial resolution of 
10~$\mu m$.

These two SiPM layers can also be used for detection of the Cherenkov 
lights when the particle passes through a dielectric medium.  
The fast timing and the RICH detector could provide particle identification with at least 3~$\sigma$ separation
between $K^{\pm}$ and $\pi^{\pm}$ up to 30~GeV/c, which covers most 
particles from Higgs decay as shown in Fig.~\ref{fig:trackpt}. 

The momentum reach of a  RICH detector~\cite{rich} depends on the dielectric medium 
and the photon detection spatial resolution. In order
to have Kaon and Pion separation of 3~$\sigma$, we need to have two type of dielectric medium for RICH:
\begin{itemize} 
\item First is an aerogel crystal bar with a refrax index of 1.025 and 2 cm 
thick, located at the radius of 1.2~m, such that the Cherenkov lights can be 
detected by the SiPM layer at a radius of 1.4~m. The extra 
material from Aerogel is about 1.5\% of the radiation length. 
The  momentum range coverage is $5<P_T<15$~GeV/c. 
\item Second is a $C4F10$ radiator with a reflax index of 1.0014, a pressure of 0.01 
atomsphere, and 30 cm think at a radius of 1.7~m. The Cherenkov 
photons are focused and reflexed by a set of mirrors to a seond SiPM layer, at the 
radius of 1.7~m. The additional material length is about 2\% of the radiation 
length including both the gas and the mirror. 
The momentum coverage is $15<P_T<30$~GeV/c. 
\end{itemize}

\begin{figure}[htpb]
\centerline{
\includegraphics[width=7.6cm]{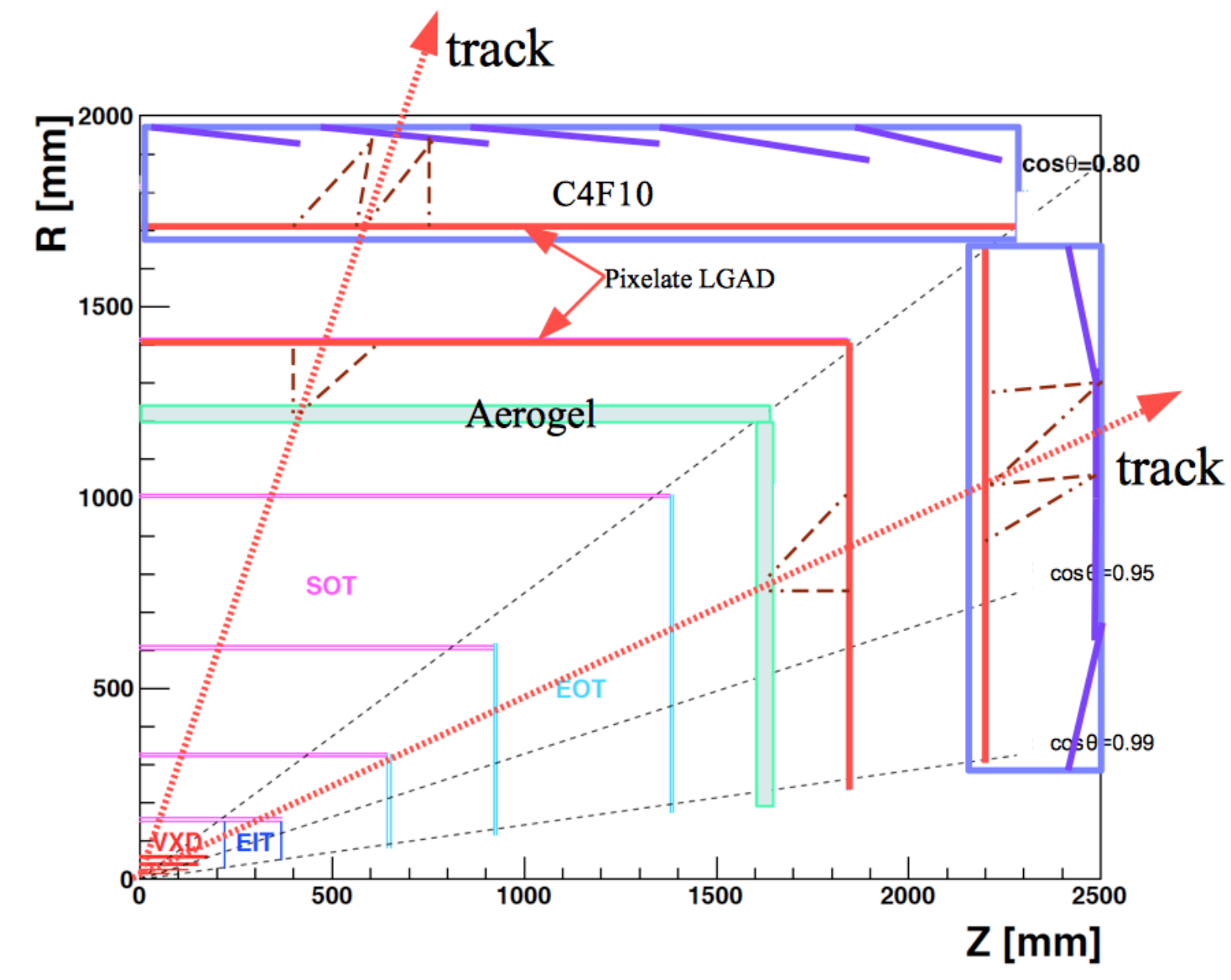}}
\caption{
A possible schematic for a full silicon tracker with TOF and RICH options.\label{fig:richp}}
\end{figure}

 A possible configuration for the full silicon tracker with fast timing and RICH
 detector option is shown in Fig.~\ref{fig:richp} for both in the barrel and endcap regions.  
 The RICH adds extra material with 5\% of the radiation length at the tracking volume.
 There might be a way to reduce the total material budget by replacing the rest of the double-sized strip layers 
 with the CMOS pixelate layers once the technology is available and affordable.                           

However the timing detector still requires signficant
R\&D efforts in order to detect both the charged track and Cherenkov light at the same time. 
There are other options possible, for example, using two types of detection separately for 
the charged tracks and the Cherenkov light.  
Work is ongoing to access various options using detailed detector simulation.       
 
\section{Conclusion}

Particle Identification (PID) plays a key role in heavy flavor physics
in high-energy physics experiments. However, its impact on the Higgs physics
is still not clear. We have explored some of 
potential of PID to improve the identification of heavy-flavour jets using identified charged Kaons 
in addition to the traditional vertexing information, which could result in improving the 
Higgs Yukawa coupling at the future e+e- collider.

\section*{Acknowledgments}

We would like to thank Tao Liu and the conference organizers for their generosity in organizing such a 
wonderful and stimulating conference. In particular, we would like to thank M. Caccia, J. Guimaraes da Costa, 
Y. Gao, S.-C. Hsu, G. Li, Z. Liang, and M. Ruan for their useful discussions.  
Work is supported in part by the Office of Science, Office of High Energy Physics, of 
the U.S. Department of Energy under contract DE-AC02-05CH11231.



\begin{thebibliography}{0}    

\bibitem{hatlas} ATLAS Collab., {\it Phys. Lett. B} {\bf 716} (2012) 1 

\bibitem{hcms} CMS Collab., {\it Phys. Lett. B} {\bf 716} (2012) 30 

\bibitem{ilc} T. Behnke {\it et al.}, {\it The International Linear Collider Technical Design Report}, arXiv:1306.6327.
 
\bibitem{cepc} CEPC Study Group, {\it CEPC Conceptual Design Report: Volume 2 - Physics and Detector}, arXiv:1811.10545.

\bibitem{fccee} M. Benedikt {\it et al.}, {\it Future Circular Collider: Vol. 2 The Lepton Collider (FCC-ee)}, CERN-ACC-2018-0057.

\bibitem{cepcsf} M. Ruan {\it et al.}, {\it Eur. J. Phys. C} {\bf 78}, 426 (2018). 
 
\bibitem{lgad} G. Pellegrini {\it et al.}, {\it Nucl. Instrum. Methods A}, {\bf 765} (2014 12.

\bibitem{ufsd} H.F.-W. Sadrozinsky {\it et al.}, {\it Nucl. Instrum. Methods A} {\bf 730} (2013) 226.

\bibitem{rich} J. Seguinot and T. Ypsilantis {\it Nucl. Instrum. Methods A} {\bf 343} (1994) 1.
 
\bibitem{hgtd} H.F.-W. Sadrozinsky {\it et al.}, {\it Nucl. Instrum. Methods A} {\bf 831} (2016) 18.

\end{thebibliography}
\end{document}